\documentclass[11pt]{article}  
\usepackage{latexsym}
\usepackage{graphicx}
\newcommand{\vev}{{\it vev}}

\newcommand{\myref}[1]{(\ref{#1})}
\def\bea{\begin{eqnarray}}
\def\eea{\end{eqnarray}}
\newcommand{\gappeq}{\mathrel{\rlap {\raise.5ex\hbox{$>$}}
{\lower.5ex\hbox{$\sim$}}}}
\newcommand{\lappeq}{\mathrel{\rlap{\raise.5ex\hbox{$<$}}
{\lower.5ex\hbox{$\sim$}}}}
\def\STr{{\rm STr}}
\def\mG{m_{\frac{3}{2}}}
\def\mg{m_{\frac{1}{2}}}
\def\half{{1\over2}}
\newcommand{\lang}{\left\langle}
\newcommand{\rang}{\right\rangle}
\newcommand{\myvev}[1]{{\lang #1 \rang}}
\newcommand{\jbar}{\bar{\jmath}}

\def\thefootnote{\fnsymbol{footnote}}

\def\Tr{{\rm Tr}}
\def\im{{\rm Im}}
\def\re{{\rm Re}}

\def\del{\delta}
\def\ux{$U(1)_X$}
\def\ua{$U(1)_a$}

\def\uo{$U(1)$}
\def\s{\bar{s}}
\def\S{\bar{S}}
\def\t{\bar{t}}
\def\[{\left [}
\def\]{\right ]}
\def\({\left (}
\def\){\right )}
\def\bl{\bar{\lambda}}

\def\pp{\partial}

\def\D{{\cal D}}
\def\G{{\cal G}}

\def\m{\bar{m}}
\def\z{\bar{z}}

\newcommand{\pl}{{\it Phys.\ Lett. }}
\newcommand{\np}{{\it Nucl.\ Phys. }}

\def\L{{\cal L}}
\def\l{\left.}
\def\r{\right|}

\def\Gev{{\rm GeV}}
\newcommand{\beq}{\begin{equation}}
\newcommand{\eeq}{\end{equation}}
\newcommand{\bear}{\begin{eqnarray}}
\newcommand{\eear}{\end{eqnarray}}

\begin{document}
\begin{titlepage}

      \hfill  LBNL-57549

      \hfill  UCB-PTH-05/13

      \hfill hep-ph/yymmmnnn

\hfill May 2005
\\[.2in]

\begin{center}

{\large \bf Effective Supergravity from the Weakly Coupled Heterotic
String}\footnote{Talk presented at the Symposium in honor of Julius
Wess, Jan. 10--11, 2004, to be published in the
proceedings.}\footnote{This work was supported in part by the
Director, Office of Energy Research, Office of High Energy and Nuclear
Physics, Division of High Energy Physics of the U.S. Department of
Energy under Contract DE-AC03-76SF00098 and in part by the National
Science Foundation under grant PHY-0098840.}

Mary K. Gaillard \\[.05in]

{\em Department of Physics,University of California, and Theoretical
 Physics Group, 50A-5101, Lawrence Berkeley National Laboratory,
 Berkeley, CA 94720, USA}\\[.2in]
\end{center}

\begin{abstract} The motivation for Calabi-Yau-like compactifications
of the weakly coupled $E_8\otimes E_8$ heterotic string theory, its
particle spectrum and the issue of dilaton stabilization are briefly
reviewed.  Modular invariant models for hidden sector condensation and
supersymmetry breaking are described at the quantum level of the
effective field theory.  Their phenomenological and cosmological
implications, including a possible origin for R-parity, are
discussed.

\end{abstract}
\end{titlepage}

\newpage
\renewcommand{\thepage}{\roman{page}}
\setcounter{page}{2}
\mbox{ }

\vskip 1in

\begin{center}
{\bf Disclaimer}
\end{center}

\vskip .2in

\begin{scriptsize}
\begin{quotation}
This document was prepared as an account of work sponsored by the United
States Government.  While this document is believed to contain
correct information, neither the United States Government nor any agency
thereof, nor The Regents of the University of California, nor any of their
employees, makes any warranty, express or implied, or assumes any legal
liability or responsibility for the accuracy, completeness, or usefulness
of any information, apparatus, product, or process disclosed, or represents
that its use would not infringe privately owned rights.  Reference herein
to any specific commercial products process, or service by its trade name,
trademark, manufacturer, or otherwise, does not necessarily constitute or
imply its endorsement, recommendation, or favoring by the United States
Government or any agency thereof, or The Regents of the University of
California.  The views and opinions of authors expressed herein do not
necessarily state or reflect those of the United States Government or any
agency thereof, or The Regents of the University of California.
\end{quotation}
\vfill

\end{scriptsize}

\vskip 2in

\begin{center}
\begin{small}
{\it Lawrence Berkeley Laboratory is an equal opportunity employer.}
\end{small}
\end{center}

\newpage
\renewcommand{\theequation}{\arabic{section}.\arabic{equation}}
\renewcommand{\thepage}{\arabic{page}}
\setcounter{page}{1}
\def\thefootnote{\arabic{footnote}}
\setcounter{footnote}{0}

\section{The weakly coupled $E_8\otimes E_8$ Heterotic String}

\subsection{Bottom up: the case for supergravity}
The primary phenomenological motivation for supersymmetry (SUSY) is
the observed large hierarchy between the $Z$ mass, characteristic of
the scale of electroweak symmetry breaking, and the reduced Planck
scale $m_P$:
$$ m_Z\approx 90\Gev \ll m_P = \sqrt{8\pi\over
G_N}\approx2\times10^{18}\Gev. $$ 
This hierarchy can be technically understood in the context of SUSY.
The conjunction of SUSY and general relativity (GR) implies
supergravity (SUGRA).  The absence of observed SUSY partners
(sparticles) requires broken SUSY in the vacuum, and the observed
particle spectrum constrains the mechanism of SUSY-breaking in the
observable sector: spontaneous SUSY-breaking is not viable, leaving
soft SUSY-breaking as the only option that preserves the technical
SUSY solution to the hierarchy problem.  This means introducing
SUSY-breaking operators of dimension three or less--such as gauge
invariant masses--into the Lagrangian for the SUSY extension of the
Standard Model (SM).  The unattractiveness of these {\it ad hoc} soft
terms suggests they arise from spontaneous SUSY breaking in a ``hidden
sector'' of the underlying theory.  Based on the above facts, a number
of standard scenarios have emerged.  These include: {\it i}) Gravity
mediated SUSY-breaking, usually understood as ``minimal SUGRA''
(mSUGRA)~\cite{can}. This scenario is typically characterized
by\newline $ m_{\rm scalars}= m_{0} \sim m_{\rm gravitino} =
m_{3\over2}> m_{\rm gauginos} = m_{1\over2}$ at the weak scale.  {\it
ii}) Anomaly mediated SUSY-breaking~\cite{hit,rs}, in which $m_0 =
m_{1\over2}=0$ classically; these models are characterized by $
m_{3\over2} >> m_0 ,\; m_{1\over2},$ and typically $m_0 >
m_{1\over2}$.  An exception is the Randall-Sundrum (RS) ``separable
potential'', constructed~\cite{rs} to mimic SUSY-breaking on a brane
spatially separated from our own in a fifth dimension; in this
scenario $m_0^2 < 0$ and $m_0$ arises first at two loops. In general,
the scalar masses at one loop depend on the details of Planck-scale
physics~\cite{bgn}. {\it iii}) Gauge mediated SUSY uses a hidden
sector that has renormalizable gauge interactions with the SM
particles, and is typically characterized by small $m_{1\over2}$.

\subsection{Top down: the case for superstring theory} 
Here the driving motivation is that superstring theory is at present
the only known candidate for reconciling GR with quantum
mechanics. These theories are consistent in ten dimensions; in recent
years it was discovered that all the consistent~\cite{gs} superstring
theories are related to one another by dualities, namely S-duality:
$\alpha\to 1/\alpha,$ and T-duality: $R\to 1/ R,$ where $\alpha$ is
the fine structure constant of the gauge group(s) at the string scale,
and $R$ is a radius of compactification from dimension D to dimension
${\rm D} -1$. These theories, as well as \mbox{D $=11$} SUGRA, are now
understood as particular limits of M-theory.  Recently, there has
been considerable activity in type I and II theories, or more
generally in theories with branes.  Similarly, the Ho\v rava-Witten
(HW) scenario~\cite{hw} and its inspirations have received
considerable attention.  Compactification of one of the 11 dimensions
of M-theory gives the HW scenario with two 10-D branes, each having an
$E_8$ gauge group.  As the length of the 11th dimension is shrunk to
zero, the two branes coincide, and the WCHS scenario~\cite{wchs} is
recovered.  This is the scenario addressed here.

\subsection{Calabi-Yau (like) compactification}
The zero-slope (infinite string tension) limit of heterotic string
theory is ten dimensional supergravity coupled to a supersymmetric
Yang-Mills theory with an $E_8\otimes E_8$ gauge group.  To make
contact with the real world, six of these ten dimensions must be
compact and here are assumed to be of order $m_P\sim10^{-32}$cm.  If
the topology of the extra dimensions were a six-torus, which has a
flat geometry, the 8-component spinorial parameter of $N=1$
supergravity in ten dimensions would appear as the four two-component
parameters of $N=4$ supergravity in four dimensions.  A Calabi-Yau
(CY) manifold leaves only one of these spinors invariant under
parallel transport; the group of transformations under parallel
transport (holonomy group) is the $SU(3)$ subgroup of the maximal
$SU(4) \cong SO(6)$ holonomy group of a six dimensional compact space.
This breaks $N=4$ supersymmetry to $N=1$ in four dimensions.  The only
phenomenologically viable supersymmetric theory at low energies is
$N=1$, because it is the only one that admits complex representations
of the gauge group that are needed to describe quarks and leptons. For
this solution, the classical equations of motion impose the
identification of the affine connection of general coordinate
transformations on the compact space (described by three complex
dimensions) with the gauge connection of an $SU(3)$ subgroup of one of
the $E_8$'s: $E_8\ni E_6\otimes SU(3)$, resulting in $E_6\otimes E_8$
as the gauge group in four dimensions~\cite{cy}.  Since the early
1980's, $E_6$ has been considered the largest group that is a
phenomenologically viable candidate for a Grand Unified Theory (GUT)
of the SM.  Hence $E_6$ is identified as the gauge group of the
``observable sector'', and the additional $E_8$ is attributed to a
``hidden sector'', that interacts with the former only with
gravitational strength couplings.  Orbifolds, which are flat spaces
except for points of infinite curvature, are more easily studied than
CY manifolds, and orbifold compactifications that closely mimic CY
compactification, and that yield realistic spectra with just three
generations of quarks and leptons, have been
found~\cite{iban,fiqs}. In this case the surviving gauge group is
$E_6\otimes\G_o \otimes E_8,\;\G_o\in SU(3)$.  The low energy
effective field theory is determined by the massless spectrum, {\it
i.e.}, the spectrum of states with masses very small compared with the
string tension and compactification scale. Massless bosons have zero
triality under an $SU(3)$ which is the diagonal of the $SU(3)$
holonomy group and the (broken) $SU(3)$ subgroup of one $E_8$.  The
ten-vectors $A_M,\; M = 0,1,\ldots 9,$ appear in four dimensions as
four-vectors $A_\mu,\;\mu = M = 0,1,\ldots 3$, and as scalars $A_m,\;
m = M-3 = 1,\cdots 6.$ Under the decomposition $E_8\ni E_6\otimes
SU(3)$, the $E_8$ adjoint contains the adjoints of $E_6$ and $SU(3)$,
and the representation ${\bf(27,3)} +
{\bf(\overline{27},\overline{3})}$.  Thus the massless spectrum
includes gauge fields in the adjoint representation of
$E_6\otimes\G_o\otimes E_8$ with zero triality under both $SU(3)$'s,
and scalar fields in ${\bf 27 + \overline{27}}$ of $E_6$, with
triality $\pm1$ under both $SU(3)$'s, together with their fermionic
superpartners.  The number of ${\bf 27}$ and ${\bf\overline{27}}$
chiral supermultiplets that are massless depends on the topology of
the compact manifold.  The important point for phenomenology is the
decomposition under $E_6\to SO(10)\to SU(5)$:
\beq\({\bf 27}\)_{E_6} =
\({\bf 16 + 10 + 1}\)_{SO(10)} = \({\bf \{\bar{5} + 10 + 1\} + \{5 +
\bar{5}\} + 1}\)_{SU(5)}.\eeq 
A ${\bf \overline{5} + 10 + 1}$ contains
one generation of quarks and leptons of the SM, a right-handed
neutrino and their scalar superpartners; a ${\bf 5 + \overline{5}}$
contains the two Higgs doublets needed in the supersymmetric extension
of the SM and their fermion superpartners, as well as color-triplet
supermultiplets. While all the states of the SM and its minimal
supersymmetric extension are present, there are no scalar particles in
the adjoint representation of the gauge group. In conventional models
for grand unification, these (or other large representations) are
needed to break the GUT group to the SM.  In string theory, this
symmetry breaking can be achieved by the Hosotani or ``Wilson line'',
mechanism in which gauge flux is trapped around ``holes'' or
``tubes'' in the compact manifold, in a manner reminiscent of the
Arahonov-Bohm effect.  The vacuum value of the trapped flux $<\int
d\ell^m A_m>$ has the same effect as an adjoint Higgs, without the
difficulties of constructing a potential for large Higgs
representations that actually reproduces the observed vacuum. When
this effect is included, the gauge group in four dimensions is \bear
&&\G_{obs}\otimes\G_{hid},
\quad\G_{obs}=\G_{SM}\otimes\G'\otimes\G_o,\quad \G_{SM}\otimes\G'\in
E_6, \quad \G_o\in SU(3),\nonumber \\ && \G_{hid}\in E_8,\quad \G_{SM}
= SU(3)_c\otimes SU(2)_L\otimes U(1)_w.
\label{eq:group}\eear

There are many other four dimensional string vacua in addition to
those described above. The attractiveness of the above picture is that
the requirement of $N=1$ SUSY naturally results in a
phenomenologically viable gauge group and particle spectrum, and the
gauge symmetry can be broken to a product group embedding the SM
without introducing large Higgs representations.  

\subsection{Gaugino Condensation and the Runaway Dilaton}
The $E_8\otimes E_8$ string theory provides a hidden sector
needed for spontaneous SUSY-breaking.  Specifically, if some subgroup
$\G_c$ of $\G_{hid}$ is asymptotically free, with a $\beta$-function
coefficient $b_c>b_{SU(3)}$, defined by the renormalization group
equation (RGE)
\beq \mu{\pp g_c(\mu)\over\pp\mu} = -b_cg_c^3(\mu) +
O(g_c^5)\label{eq:rge},\eeq 
confinement and fermion condensation will occur at a scale
$\Lambda_c\gg\Lambda_{QCD}$, and hidden sector gaugino condensation
$<\bl\lambda>_{\G_c} \ne 0,$ may induce~\cite{nilles} supersymmetry
breaking.  To discuss supersymmetry breaking in more detail, we need
the low energy spectrum resulting from the ten-dimensional gravity
supermultiplet that consists of the 10-D metric $g_{MN}$, an
antisymmetric tensor $b_{MN}$, the dilaton $\phi$, the gravitino
$\psi_M$ and the dilatino $\chi$.  For the class of CY and orbifold
compactifications described above, the massless bosons in four
dimensions are the 4-D metric $g_{\mu\nu}$, the antisymmetric tensor
$b_{\mu\nu}$, the dilaton $\phi$, and certain components of the
tensors $g_{mn}$ and $b_{mn}$ that form the real and imaginary parts,
respectively, of complex scalars known as K\"ahler moduli.  The number
of moduli is related to the number of particle generations (\# of
${\bf 27}$'s $-$ \# of ${\bf\overline{27}}$'s).  In three generation
orbifold models there are at least three moduli $t_I$ whose $vev$'s
$<{\rm Re}t_I>$ determine the radii of the three tori of the compact
space.  They form chiral multiplets with fermions $\chi^t_I$ obtained
from components of $\psi_m$.  The 4-D dilatino $\chi$ forms a chiral
multiplet with with a complex scalar field $s$ whose $vev$ $ <s> =
g^{-2} - i\theta/8\pi^2$ determines the gauge coupling constant and
the $\theta$ parameter of the 4-D Yang-Mills theory.  The
``universal'' axion Im$s$ is obtained by a duality transformation from
the antisymmetric tensor $b_{\mu\nu}$: $\pp_\mu{\rm
Im}s\leftrightarrow \epsilon_{\mu\nu\rho\sigma}\pp^\nu
b^{\rho\sigma}.$ Because the dilaton couples to the (observable and
hidden) Yang-Mills sector, gaugino condensation induces a
superpotential for the dilaton superfield\footnote{Throughout I use
capital Greek or Roman letters to denote a chiral superfield, and the
corresponding lower case letter to denote its scalar component.} $S$:
\beq W(S)\propto e^{-S/b_c}.\label{eq:dil}\eeq The vacuum value $
<W(S)> \propto \left<e^{-S/b_c}\right> = e^{-g^{-2}/b_c}= \Lambda_c$
is governed by the condensation scale $\Lambda_c$ as determined by the
RGE (\ref{eq:rge}).  If it is nonzero, the gravitino acquires a mass
$m_{3\over2}\propto<W>$, and local supersymmetry is broken.

The superpotential (\ref{eq:dil}) results in a potential for the
dilaton of the form $ V(s)\propto e^{-2\re s/b_c},$ which has its
minimum at vanishing vacuum energy and vanishing gauge coupling: $<\re
s> \to\infty,\; g^2\to 0$.  This is the notorious runaway dilaton
problem.  The effective potential for $s$ is in fact determined from
anomaly matching: $\delta\L_{eff}(s,u) \longleftrightarrow
\delta\L_{hid}({\rm gauge}),$ where $u, \;\left<u\right> = \left<
\bl\lambda\right>_{\G_c},$ is the lightest scalar bound state of the
strongly interacting, confined gauge sector.  Just as in QCD, the
effective low energy theory of bound states must reflect both the
symmetries and the anomalies, {\it i.e.} the quantum induced breaking
of classical symmetries, of the underlying Yang-Mills theory.  It
turns out that the effective quantum field theory (QFT) is anomalous
under T-duality.  Since this is an exact symmetry of heterotic string
perturbation theory, it means that the effective QFT is incomplete.
This is cured by including model dependent string-loop threshold
corrections~\cite{thresh} as well as a ``Green-Schwarz'' (GS)
counter-term~\cite{gs1}, analogous to the GS mechanism in 10-D
SUGRA.  This introduces dilaton-moduli mixing, and the gauge coupling
constant is now identified as
\beq g^2= 2\left<\ell\right>,\quad\ell^{-1} = 2\re s -
b{\sum_I} \ln(2\re t^I),\label{dual}\eeq
where $b\le b_{E_8} = 30/8\pi^2$ is the coefficient of the GS term,
and and $\ell$ is the scalar component of a linear superfield $L$ that
includes the two-form $b_{\mu\nu}$ and is dual to the chiral
superfield $S$ in the supersymmetric version of the
two-form/axion duality mentioned above.  The GS term introduces a
second runaway direction, this time at strong coupling: $V \to -
\infty$ for $g^2\to\infty$. The small coupling behavior is unaffected,
but the potential becomes negative for $\alpha = \ell/2\pi >
.57$. This is the strong coupling regime, and nonperturbative string
effects cannot be neglected; they are expected~\cite{shenk} to modify
the K\"ahler potential for the dilaton, and therefore the potential
$V(\ell,u)$.  It has been shown~\cite{bgw1} that these contributions
can indeed stabilize the dilaton.  Retaining just one or two terms of
the suggested parameterizations~\cite{shenk,bd1} of the nonperturbative
string corrections: $a_n\ell^{-n/2}e^{-c_n/\sqrt{\ell}}$ or
$a_n\ell^{-n}e^{-c_n/\ell},$ the potential can be made
positive-definite everywhere and the parameters $a_n,c_n$ can be
chosen to fit two data points: the coupling constant $g^2\approx 1/2$
and the cosmological constant $\Lambda \simeq 0$.  This is fine
tuning, but it can be done with plausible (order 1) values for the
parameters $c_n, a_n$.  If there are several condensates with
different $\beta$-functions, the potential is dominated by the
condensate with the largest $\beta$-function coefficient $b_+$, and
the result is essentially the same as in the single condensate case,
except that a small mass is generated for the axion $a=\im s$.
In these models the presence of $\beta$-function coefficients generate
mass hierarchies that have interesting implications for cosmology
and the spectrum of sparticles--the supersymmetric partners of the SM
particles.

\section{Modular invariant gaugino condensation}
In this section I will summarize results~\cite{bgw,ggm} from the study
of modular (T-duality) invariant effective Lagrangians for gaugino
condensation.  These are characterized in particular by
\begin{itemize}
\item Dilaton dominated supersymmetry breaking. The auxiliary fields
of the T-moduli (or K\"ahler moduli) have vanishing vacuum values
(\vev's): $\lang F^T\rang = 0,$ thus avoiding a potentially dangerous
source of flavor changing neutral currents (FCNC).
\item The constraint of vanishing (or nearly so) vacuum energy leads
to a variety of mass hierarchies that involve the $\beta$-function
coefficient of the condensing gauge group.
\end{itemize}

One starts above the (reduced) Planck scale $m_P$ with the heterotic
string theory in 10 dimensions.  Just below the string scale $\mu_s =
g_s m_P$, where $g_s$ is the gauge coupling at the string scale,
physics is described by $N=1$ modular invariant supergravity in four
dimensions, where here modular invariance refers to T-duality under
which the K\"ahler moduli $T$ transform as
\beq T\to\frac{a T - i b T}{i c T + d},\qquad a d - b c = 1, \qquad
a,b,d,c\in \mathcal{Z}.\label{mdtr}\eeq
Modular invariance -- and in many compactifications~\cite{jg} a \uo\,
gauge group factor called \ux\, -- is broken by anomalies at the
quantum level of the effective field theory, and the symmetry is
restored by an appropriate combination of threshold
effects~\cite{thresh} and four dimensional GS
term(s)~\cite{gs1,gs2}.  The precise form of these loop effects in the
Yang-Mills sector of the effective supergravity theory have been
determined by matching the string and field theory amplitudes at the
quantum level~\cite{gt}.

If an anomalous \uo\, is present, the corresponding GS term leads to a
Fayet-Illiopoulos (FI) D-term in the effective Lagrangian~\cite{gs2}
and some \uo-charged scalars $\phi^A$ acquire \vev's at a scale
$\mu_D$ one or two orders of magnitude below the Planck scale such
that the overall D-terms vanish:
\beq \lang\frac{1}{\ell(s,\s)}\sum_A q^a_A\prod_I(t^I +
\t^I)^{n_I^A}|\phi^A|^2\rang = \half\delta_X\delta_{Xa},\label{dterm}\eeq
where $\delta_X\ell$ is the coefficient of the FI term, $n^A$ is the
modular weight of $\phi^A$, $q^a_A$ is its charge under the gauge
group factor \ua, and $t,s$ are the scalar components of the K\"ahler
moduli and dilaton chiral superfields $T,S$. The function $\ell(s,\s)
= \ell(s + \s)$ is the dilaton field in the dual, linear
supermultiplet formulation; in the classical limit $\ell = (s +
\s)^{-1}$.  The combination of fields that gets a vacuum value is
modular invariant.  Thus modular invariance, as well as local
supersymmetry, is unbroken at this scale, and the moduli fields $s,t$
remain undetermined~\cite{ggu}. The $\phi^A$ vacuum is generically
characterized by a high degree of further degeneracy~\cite{bdfs}
that may lead to problems for cosmology.

At a lower scale $\mu_c$, a gauge group $\mathcal{G}_c$ in the hidden
sector becomes strongly coupled, and gauginos as well
$\mathcal{G}_c$-charged matter condense.  The potential generated for
the moduli is T-duality invariant and the K\"ahler moduli $T$ are
stabilized at self-dual points with $\lang F^T\rang = 0$, while $\lang
F^S\rang\ne 0$, so that, in the absence of an anomalous \uo,
supersymmetry breaking is dilaton mediated~\cite{bgw}. In the presence
of an anomalous \uo, \vev's of D-terms are generically generated as
well and tend to dominate supersymmetry breaking; these may be
problematic for phenomenology.  On the plus side, at least some of the
degeneracy of the $\phi^A$ vacuum is lifted by $\phi^A$ couplings to
the condensates~\cite{ggm}.
  
To briefly summarize the phenomenology of these models, the condition
of vanishing vacuum energy introduces the $\beta$-function coefficient
of the condensing gauge group $\mathcal{G}_c$
into the supersymmetry breaking parameters in
such a way as to generate a variety of mass hierarchies.  Defining
\beq b_c = \frac{1}{16\pi^2}\(3C^c - C^c_M\),\eeq
where $C^c(C^c_M)$ is the adjoint (matter) quadratic Casimir for
$\mathcal{G}_c$, in the absence of an anomalous \uo\, one has at the
condensation scale~\cite{bgw} (one can also have $m_0\sim m_T\gg\mG$ if
gauge-charged matter couples to the GS term)
\bea m_0 &=& \mG,\qquad m^a_{\half} =
\frac{4b^2_c}{9}g_a(\mu_c)\mG,\nonumber\\ m_T
&\approx&\frac{b}{b_c}\mG, \qquad m_S\sim b_c^{-2}\mG,\qquad m_a =0.
\label{masses}\eea
where $m_{0,\half,\frac{3}{2}}$ refer to observable sector scalars and
gauginos, and the gravitino, respectively; $m_{T,S,a}$ are the
K\"ahler moduli, dilaton and universal axion masses.  The expression
for $m_T$ assumes $b\gg b_c$, where $b$ is the $\beta$-function
coefficient appearing in the modular invariance restoring GS
term~\cite{gs1}. For example in the absence of Wilson lines, $b= b_{E_8}
\approx .57$, and viable scenarios for electroweak symmetry
breaking~\cite{gnph} and for neutralinos as dark matter~\cite{bhn}
require $b_c\approx .05-.06$.  These numbers give desirably large
moduli and dilaton masses, while the scalar/gaugino mass ratio is
perhaps uncomfortably large, but no worse than in many other models.

When Wilson lines are present the condition $b\gg b_c$ may not hold;
for example $b_c = b$ in a $Z_3$ compactification~\cite{fiqs} with an
$SO(10)$ hidden sector gauge group; this would give vanishing T-moduli
masses in the above class of models.  However when an anomalous \uo\,
is present, the T-moduli couplings to the condensates are modified,
giving additional contributions to their masses, and a hierarchy with
respect to the gravitino mass can still be maintained~\cite{ggm}. In
this scenario the gaugino, dilaton and axion masses are determined
only by the dilaton potential, as before.  A stable vacuum with a
positive metric for the dilaton is most easily achieved in a
``minimal'' class of models in which the number of Standard Model (SM)
gauge singlets that get \vev's at the scale $\mu_D$ is equal to the
number $m$ of broken \uo's (in which case there are no massless
``D-moduli''~\cite{bdfs} associated with the degeneracy of the
\uo-charged $\phi^A$ vacuum), or $N$ replicas of these with identical
\uo\, charges [yielding $(N-1)m$ D-moduli].  In this case the gaugino,
dilaton and axion\footnote{The possibility that an axion mass may be
generated by higher dimension operators~\cite{bd1} is under
study~\cite{butter}.}  masses are unchanged from (\ref{masses}).  The
most significant change from the above scenario is a D-term
contribution to scalar squared masses $m_0^2$ that is proportional to
their \uo\, charges. At weak coupling, and neglecting nonperturbative
effects, this term dominates the one in (\ref{masses}) by a factor
$b_c^{-2}\gg1$, and is not positive semi-definite.  Thus unless SM
particles are uncharged under the broken \uo's (or have charges that,
in a well-defined sense~\cite{ggm}, are orthogonal to those of the
$\phi^A$ with large \vev's), these models are seriously challenged by
the SM data: a very high scalar/gaugino mass ratio for positive
$m_0^2$, and the danger of color and electromagnetic charge breaking
if $m_0^2<0$.

\section{QFT quantum corrections}

The above results were obtained at tree level in the effective
supergravity theory for gaugino condensation, which includes QFT and
string quantum corrections to the strongly coupled gauge sector whose
elementary degrees of freedom have been integrated out, as well as the
four dimensional Green-Schwarz (GS) terms needed at the quantum level
to cancel field theory anomalies.  In addition, the logarithmically
divergent and finite (``anomaly mediated''~\cite{hit,rs,anom})
one-loop corrections to soft supersymmetry-breaking parameters have
been extensively studied~\cite{gnw,bgn}. These analyses did not
include quadratically divergent loop corrections which are
proportional to terms in the tree Lagrangian, and are suppressed by
the loop expansion parameter
\beq \epsilon = 1/16\pi^2.\label{ep}\eeq
However, since some of these terms have coefficients proportional to
the number of fields in the effective supergravity theory, it has been
argued that they may not be negligible.  In particular, their
contributions to the cosmological constant~\cite{ckn} and to flavor
changing neutral currents~\cite{clm} have been emphasized. 
Both are important for the phenomenology of the above
condensation models; thus we need to revisit~\cite{gnq} their effects.

When local supersymmetry is broken, there is a quadratically
divergent one-loop contribution to the vacuum energy~\cite{zum}
\beq \lang V_{\rm 1-loop} \rang \ni \frac{\Lambda^2}{32\pi^2}
\lang\STr \mathcal{M}^2\rang,\eeq
where $\mathcal{M}$ is the field-dependent mass matrix, and the
gravitino contribution is gauge dependent.  For example in minimal
supergravity~\cite{can} with $N_{\chi}$ chiral and $N_G$
Yang-Mills  superfields, one obtains, using the gravitino gauge fixing
procedure of Ref~\cite{gjs}.
\begin{equation} 
\lang \delta V_{\rm 1-loop} \rang
\ni\frac{\Lambda^2}{16\pi^2}\(N_{\chi} m_0^2 -N_G\mg^2 + 2\mG^2\) .
\label{STr} \end{equation}
In the MSSM we have $N_{\chi} = 49$ and $N_G = 12$. The much larger
field content of a typical $Z_3$ orbifold
compactification~\cite{cmm,jg} of the $E_8\otimes E_8$ heterotic string
has $N_\chi\gappeq 300$ and $N_G \lappeq 65$, suggesting~\cite{ckn}
that this contribution to the vacuum energy is always positive.

However, in order to maintain manifest supersymmetry, a supersymmetric
regularization of ultraviolet divergences must be used.  Pauli-Villars
(PV) regularization~\cite{pv} meets this criterion.  The regulation of
quadratic divergences requires {\it a priori} two subtractions; in the
context of PV regularization, the number $S$ of subtractions is the
number of PV fields for each light field. Once the divergences are
regulated ({\it i.e.} eliminated), we are left with the replacement
\begin{equation}
\Lambda^2 \STr\mathcal{M}^2 \to \STr \mu^2 \mathcal{M}^2 \ln(\mu^2)
\eta_S, \qquad \eta_S = \sum_{q=1}^S \eta_q \lambda_q \ln \lambda_q,
\label{subtract} \end{equation}
where $\mu$ represents the scale of new physics, and the parameter
$\eta_S$ reflects the uncertainty in the threshold for the onset of
this new physics. The squared PV mass of the chiral supermultiplet
$\Phi^q$ is $\lambda_q\mu^2$ (so $\lambda_q>0$), and $\eta_q = \pm1$
is the corresponding PV signature.  The sign of the effective cut-off
is determined by the sign of $\eta_S$, which is positive definite
only\footnote{See appendix C of \cite{bgsig}. and the discussion
in \cite{gnq}.} if $S\le3$. Cancellation of all the ultraviolet
divergences of a general supergravity theory requires~\cite{bgpr} at
least 5 PV chiral multiplets for every light chiral multiplet and even
more PV supermultiplets to regulate gauge loops.  Therefore one cannot
assume that the effective cut-offs are all positive.

More importantly, the Lagrangian constructed using a simple cut-off
does not respect supersymmetry.  With a supersymmetric PV
regularization, PV masses arise from quadratic couplings in the
superpotential
\beq W_{PV} \ni \mu_{I J}(Z^k)Z^I_{PV}Z^J_{PV},\qquad
\left.Z^k \right|= z^k.\label{wpv}\eeq
Then the squared cut-off in (\ref{STr}) is replaced by suitably
weighted linear combinations of PV squared masses
\beq \Lambda^2 \to (M^2)^I_J = e^K K^{I\bar K}(z)K^{\bar L M}(z)
\bar\mu_{\bar K\bar L}(\z)\mu_{M J}(z)\eeq
that are generally field-dependent.  Moreover, the couplings
(\ref{wpv}) induce additional terms proportional to $M^2$ that
cannot be obtained by a straight cut-off procedure.  The resulting
effective Lagrangian takes the form~\cite{pv}
\beq \L^1_{eff} = \L_{\rm tree}(g,K) + \L_{\rm 1-loop} = \L_{\rm
tree}(g_R,K_R) + O(\epsilon\ln\Lambda_{eff}^2) + O(\epsilon^2),\eeq
where 
\beq K_R = K + \Delta K \eeq
is the renormalized superpotential.  The action obtained in this
way is only perturbatively supersymmetric:
\beq \delta S^1_{eff}  = \int d^4x \delta\L^1_{eff} = O(\epsilon^2).\eeq
Writing 
\beq \Delta K = \frac{\epsilon}{2}\[N\Lambda_\chi^2 -
4N_G\Lambda_G^2 +O(1)\Lambda^2_{\rm grav}\]
+ O(\epsilon\ln\Lambda_{eff}^2) + O(\epsilon^2),\label{delk} \eeq
where $\Lambda_{\chi,G,{\rm grav}}$ are the effective cut-offs for chiral,
gauge and gravity loops, and $\Lambda_{eff}$ is a generic effective
cutoff, if $N_\chi,N_G\sim \epsilon^{-1}$, we must retain the full
effective Lagrangian as derived from $K_R$.  This amounts to
resuming the leading terms in $\epsilon N\Lambda_{eff}^2$, with the
result, as dictated by supersymmetry, just a correction to the
K\"ahler potential.  I will discuss the consequences of this correction
in the remainder of this section.

\subsection{The vacuum energy}
Consider first the possibility that we can choose the $Z^k$-dependence
of the PV K\"ahler potential and superpotential such that the
effective cutoffs are constant.  For example, one needs PV superfields
$Z^I_{PV}$ with the same K\"ahler metric as the light superfields
$Z^i$: $K^Z_{I\bar M} = K_{i\m}.$ If we introduce superfields $Y_I$ with
K\"ahler metric: $K_Y^{I\bar M} = e^{-K}K^{-1}_{i\m} = e^{-K}K^{i\m},$
the superpotential coupling
\beq W_{PV} = \mu Z^I Y_I\eeq
yields a constant squared mass $M^2 = \mu^2$ if $\mu$ is constant,
and the quantum corrected potential just reads
\beq V_{eff} = \D + e^{\Delta K}\(F^i K_{i\m} F^{\m} - 3\mG^2\)_{\rm
tree} + O(\epsilon\ln\Lambda_{eff}^2).\label{const}\eeq
If supersymmetry breaking is F-term induced: $\lang\D\rang=0$, the
tree level condition $\lang F^i K_{i\m} F^{\m} = 3\mG^2\rang$ for
vanishing vacuum energy is unmodified by these quantum corrections.

However not all PV masses can be chosen to be constant because of the
anomaly associated with K\"ahler transformations $K(Z,\bar Z)\to
K(Z,\bar Z) + F(Z) + \bar F(\bar Z)$ that leave the classical
Lagrangian invariant.  In the presence, for example, of an anomalous
\ux, with generator $T_X$, there is a quadratically divergent term
proportional to $\Tr T_X\Lambda^2$ that cannot be canceled by
\ux-invariant PV mass terms, since the contribution to $\Tr T_X$ from
each pair in the invariant superpotential cancels. As a consequence,
there must be some PV masses $\propto e^{a V_X}$, where $V_X$ is the
\ux\, vector superfield.  Similarly, in the presence of a K\"ahler
anomaly there is a term
\beq\L_{\rm 1-loop}\ni c\epsilon K_{i\m}\D_\mu
z^i\D^\mu\z^{\m}\Lambda^2,\label{kanom}\eeq
that cannot be canceled unless some PV superfields have masses
$M^2_{PV}\propto e^{\alpha K}$.  In addition, PV regulation of
the gauge + dilaton sector requires some PV masses proportional
to the field-dependent string-scale gauge coupling constant:
$M^2_{PV}\propto g^2_s(s,\s) = 2(s + \s)^{-1}.$  

What might be the effects of this field-dependence on the condensation
models described above?  
The modular invariance of these models assures that the
moduli $T$ are stabilized at self-dual points with vanishing
auxiliary fields: $\lang F^{T}\rang =0$. Supersymmetry breaking
is dilaton-dominated and the condition for vanishing vacuum energy
at tree level in the effective theory relates $\lang F^S\rang$
to the gravitino mass which in turn constrains the dilaton
K\"aher metric:
\beq {2\over\sqrt{3}}b_c\approx.05\le \lang K^{-\half}_{S\S}\rang
\ll \l\lang K^{-\half}_{S\S}\rang\r_{\rm classical} = 2g^{-2}_s
 \approx 4,\label{vacen} \eeq
with the approximate value of $g_s$ inferred from low energy data.  It
is clear that (\ref{vacen}) cannot be satisfied without a modification
of the K\"ahler potential for the dilaton; the approach~\cite{bgw1}
used here is to invoke nonperturbative string~\cite{shenk} and/or
QFT~\cite{bd1} corrections to the dilaton K\"ahler potential.
Avoiding dangerously large D-term contributions to scalar masses in
the presence of an anomalous \uo\, may further require~\cite{ggm}
\beq -\lang K_S\rang \approx{3\over2}b_c^{-1}\approx30
\gg -\l\lang K_S\rang\r_{\rm classical} = g_s^2/2 \approx 1/4, \eeq
suggesting that weak coupling may not be viable~\cite{hw,wit,bd97}.  On
the other hand, if $\Lambda\sim C e^{\alpha K}$, with $C\alpha^n$ large
and positive, it might be possible to reinterpret part of the needed
modification of the dilaton K\"ahler potential in terms of
perturbative quantum corrections~\cite{gnq}.

\subsection{Flavor Changing Neutral Currents}
The tree potential of an effective supergravity theory includes a term
\beq V_{\rm tree}\ni e^K K_i
K_{\jbar}K^{i\jbar}|W|^2,\label{vkahl}\eeq
and the quadratically divergent one-loop corrections generate a term
\beq V_{\rm 1-loop}\ni e^K K_i K_{\jbar}R^{i\jbar}|W|^2, \qquad
R^{i\jbar} = K^{i\bar k}R_{\bar k l} K^{k\jbar}.\label{vricc}\eeq
where $R_{i\jbar}$ is the K\"ahler Ricci tensor.  The contribution
(\ref{vricc}) simply reflects the fact that the leading divergent
contribution in a nonlinear sigma model is a correction to the
K\"ahler metric proportional to the Ricci tensor (whence, {\it e.g.},
the requisite Ricci flatness of two dimensional conformal field
theories).  Since the Ricci tensor involves a sum of K\"ahler Riemann
tensor elements over all chiral degrees of freedom, a large, order
$N_\chi$, coefficient may be generated~\cite{clm}. However, the
supersymmetric completion of the potential in any given order in 
perturbation theory yields (in the absence of D-term contributions)
the scalar squared mass matrix
\beq (m^2)^i_j = \delta^i_j\mG^2 - \myvev{\tilde R}^i_{j k\m}{\tilde
F}^k {\tilde{\bar F}}^{\m},\eeq
where ${\tilde R}^i_{j k\m}$ is an element of the Riemann tensor
derived from the fully renormalized K\"ahler metric, and ${\tilde
F^i}$ is the auxiliary field for the chiral superfield $\Phi^i$,
evaluated by its equation of motion using the quantum corrected
Lagrangian.  Since the latter is perturbatively modular invariant, the
K\"ahler moduli $t^I$ are still stabilized at self-dual points with
$\myvev{\tilde F^T}=0$. Classically we have $R^A_{B S\S}=0$ where the
indices $A,B$ refer to gauge-charged fields in the observable sector.
This need not be true at the quantum level.  For example, if, as
suggested above, the quantum correction to the K\"ahler potential
includes a term
\beq \Delta K = {1\over32\pi^2}\STr\Lambda^2_{eff} \ni {c
N_\chi\over32\pi^2}e^{\alpha K},\eeq
we get
\beq \myvev{{\tilde R}^A_{B S\S}} = \delta^A_B{c N_\chi\over32\pi^2}
\alpha^2e^{\alpha K}\(K_{S\S} + \alpha K_S K_{\S}\),\eeq
which is flavor diagonal, and therefore FCNC safe.

\section{R-parity}
The self-dual vacua
\beq \myvev{t^I} = T_{s d} = 1\quad {\rm or}\quad e^{i\pi/6}\eeq
are invariant under \myref{mdtr} with
\beq b^I = - c^I = \pm 1, \qquad a^I= d^I = 0\quad {\rm or}\quad
\cases{a^I = b^I,\; d^I = 0\cr d^I = c^I,\;a^I = 0\cr} , \qquad
F^I =  n i{\pi\over2}\quad{\rm or}\quad n i{\pi\over3}\label{z2z3}.\eeq
The hidden sector condensates that get
\vev's break this further to a subgroup $G_R$ with
\beq i\im F = F = \sum_I F^I = 2n i\pi,\label{subg}\eeq
under which $\lambda_L\to e^{-{i\over2}\im F}\lambda_L = \pm
\lambda_L$.  Observable sector gauge-charged matter chiral
supermultiplets transform as
\beq \Phi^A \to e^{i\del^A + \sum_I n_I^A F^I}
\Phi^A = R(F^I,n_I^A)\Phi^A.\label{deltr}\eeq
For example a $Z_3$ orbifold has untwisted sector fields $U^{A I}$,
and twisted sector fields $T^A$ and $Y^{A I}$ with modular weights
\beq \(n_I^{A J}\)_U = \del_I^J,\qquad n_I^A =
\({2\over3},{2\over3},{2\over3}\),\qquad \(n_I^{A J}\)_Y =
\({2\over3},{2\over3},{2\over3}\) + \del_I^J,\eeq 
and moduli independent phases
\beq \del^U = 0, \qquad\del^T = - {2\over3}\del, \qquad\del^{Y^J} = -
{2\over3}\del - 4\del_I, \qquad \del = \sum\del^I,\eeq
with $\del = 2\pi n$ for the subgroup defined by \myref{subg}. If some
modular covariant fields $\phi^A$ acquire large \vev's that break some
\uo\, gauge factors near the string scale, the transformation property
\myref{deltr} can be modified to include a discrete \ua\,
transformations such that the vacuum remains invariant.  Similarly,
below the electroweak scale where the Higgs fields acquire \vev's the
residual symmetry involves a discrete transformation under the \uo$_w$
of the SM such that $R(H_u) = R(H_d) = 1$ (the presence of a
$\mu$-term requires $R(H_u)R(H_d)=1$). Then we obtain an effective
R-parity that forbids baryon and lepton number violation while
allowing other MSSM couplings provided the remaining MSSM chiral
supermultiplets have R-charges $R(\Phi^A) = R(F^I,n^A_I,q_A^a)$ that
satisfy~\cite{mkg}
\bea R(Q) &=& e^{2i\pi\beta},\qquad R(Q^c)
= e^{-2i\pi\beta}, \qquad
R(L) = e^{2i\pi\alpha},\qquad
R(L^c) = e^{-2i\pi\alpha},\label{phases}\eea
with $\beta\ne {n\over3},\; 0< \alpha,\beta< 1.$ Since these phases
need not be $\pm1$, dimension-five operators that violate baryon and
lepton number will also be forbidden provided $3\beta + \alpha\ne n$,
which is an advantage over the conventional definition of R-parity.

\section{Issues and open questions}
Other issues relevant to the viability of the WCHS are
under active investigation.  They include
\begin{itemize}
\item Can the universal axion be identified with the QCD/Peccei-Quinn
axion~\cite{butter}?
\item Will the LHC be able to distinguish the WCHS from other
scenarios~\cite{cln}?
\item Is there a specific vacuum of the WCHS such that
\begin{itemize} 
\item The $\beta$-function of the hidden sector condensing gauge
group yields viable electroweak symmetry breaking and dark matter
scenarios [{\it e.g.} $b_c\approx .05-.06$ in the absence of an
anomalous \uo]?
\item D-term contributions to squark, slepton and Higgs masses
are absent or highly suppressed?
\item The desired R-parity emerges~\cite{mkg}?
\item A see-saw mechanism for neutrino masses is present~\cite{gkn}?
\item A $\mu$-parameter of about a TeV is natural~\cite{ant}?
\item The correct Yukawa textures arise~\cite{Giedt:2000}?  
\end{itemize}
\end{itemize}
In the present context suppression of Yukawa couplings could be due to
string selection rules that allow some superpotential couplings only
in terms of very high dimension: $W\ni Q_i Q^c_j H\prod_{A = 1}^{n_{i
j}}\Phi^A$, with \vev's $\myvev{\phi^A}\sim (.1-.01)m_{\rm Planck}$
arising at the \ux-breaking scale $\Lambda_D$.

There is a complete classification of the observable~\cite{cmm} and
hidden~\cite{jg} sectors of $Z_3$ orbifolds, but only one of
these~\cite{fiqs} has been studied in detail, and it fails the above
tests. A more complete survey of heterotic string vacua could help to
determine if any scenario of the class considered here might be able
to describe nature. A more general list of interesting questions about
the relevance of string theory to the real world can be found
in~\cite{bkln}.

\section*{Acknowledgments}
I am indebted to my many collaborators.  This work was supported in
part by the Director, Office of Energy Research, Office of High Energy
and Nuclear Physics, Division of High Energy Physics of the
U.S. Department of Energy under Contract DE-AC03-76SF00098 and in part
by the National Science Foundation under grant PHY-95-14797.

\end{document}